\documentstyle[epsfig,multicol,aps]{revtex}

\newcommand{\Beginrule}{\vskip 3pt\noindent\hbox{%
\vbox{\hbox to 9cm{\hfill}}\vbox{\hrule width 9cm}} \vskip 3pt}

\def\be{\begin{equation}}
\def\ee{\end{equation}}
\def\bea{\begin{eqnarray}}
\def\eea{\end{eqnarray}}
\def\bma{\begin{mathletters}}
\def\ema{\end{mathletters}}

\begin{document}
\input{epsf.sty}
\draft

\title{Quantum repeaters based on entanglement purification}

\author{W. D\"ur$^1$, H.- J. Briegel$^{1,2,*}$, J. I. Cirac$^{1}$, 
and P. Zoller$^1$}
\address{$^{1}$Institut f\"ur Theoretische Physik, Universit\"at Innsbruck,
Technikerstrasse 25, A--6020 Innsbruck, Austria.\\
$^{2}$Departamento de Fisica, Universidad de Castilla-La Mancha,
13071 Ciudad Real, Spain.}
\date{\today}
\maketitle

\begin{abstract}
We study the use of entanglement purification for quantum communication over long 
distances. For distances much longer than the coherence length of a corresponding noisy quantum channel, the fidelity of transmission is usually so low that standard purification methods are not applicable. It is however possible to divide the channel into shorter segments that are purified separately and then connected by the method of entanglement swapping. This method can be much more efficient than schemes based on
quantum error correction, as it makes explicit use of two-way classical communication.
An important question is how the noise, introduced by imperfect local 
operations (that constitute the protocols of purification and the entanglement swapping), accumulates in such a compound channel, and how it can be kept below a certain noise level.
To treat this problem, we first study the applicability and the efficiency of entanglement purification protocols in the situation of imperfect local operations.  We then present a scheme that allows entanglement purification over arbitrary long channels and tolerates errors on the per-cent level. It requires a polynomial overhead in time, and an overhead in local resources that grows only logarithmically with the length of the channel.
\end{abstract}
\pacs{PACS: 3.67.Hk, 3.67.-a, 3.65.Bz, 42.50.-p}


\section{Introduction}

In quantum communication, qubits \cite{schumacher96} are sent across quantum channels that connect distant nodes of a quantum network. In general, these quantum channels are noisy and therefore limit the fidelity of the transmission. A bottleneck for quantum communication 
is the scaling of the probability of transmission errors with the length of the channels 
connecting the nodes. Different from classical (digital) communication, the ``signals"
in quantum communication consist of single qubits which may be entangled. Owing to fundamental principles, these qubits cannot be cloned \cite{wootters82,dieks82} nor 
amplified \cite{glauber86} without destroying the essential quantum feature.
This fact limits the maximum distance between the nodes to a few multiples of the absorption length (or the coherence length) of the channel, and poses a severe restriction on any practical application. 
For example, in recent experiments on quantum cryptography \cite{tittel97}, single photons are sent through optical fibers with a given absorption and depolarization length. This has two effects: $(i)$ to transmit a photon without absorption, the number of trials scales exponentially with $l$; $(ii)$ even when a photon arrives, the fidelity of the transmitted state decreases exponentially with $l$. In the experiment, the distance between the nodes is therefore presently limited by a few multiples of the absorption length of the fiber \cite{footnote0}.

In the context of fault-tolerant quantum computing \cite{fault_tolerant},
Knill and Laflamme \cite{knill96} have discussed an important scheme that allows, 
in principle, to transmit a qubit over arbitrarily long distances with a
polynomial overhead in the resources. Their method requires to encode a
single qubit into a concatenated quantum code (i.e.\ an entangled state of a large number of qubits) and to operate on this code repeatedly while it is propagated through the channel. The tolerable error probabilities for transmission are less than $10^{-2}$,
whereas for local operations they are less than $5\times 10^{-5}$. This
seems to be outside the range of any practical implementation in the
near future. A crucial figure for any experiment will be the number
of particles that can be manipulated locally in a coherent fashion, together 
with the precision with which such local manipulations can be realized.  
For instance, even when the number of particles scales only polynomially 
with the length of the channel, a polynomially large number may still be far too 
large for any practical implementation.

A central tool in the theory of quantum information, specifically 
for quantum communication is entanglement purification \cite{bennett96a,bennett96b,gisin96,deutsch96}.
It allows, in principle, to create maximally entangled states of particles at different
locations even if the channel that connects those locations is noisy \cite{schumacher96}.
These entangled particles can then be used for faithful teleportation \cite{bennett93} 
or secure quantum cryptography \cite{deutsch96,ekert91}. 
It seems therefore natural to use this method as an ingredient for a quantum repeater.
Furthermore, it allows highly efficient two-way protocols which can not be 
realized with quantum error correction procedures.
If a noisy quantum channel is much longer than its coherence length, one cannot
directly employ standard purification schemes. These schemes first create an ensemble of low fidelity EPR pairs across the channel, and then purify/distill a few perfect EPR pairs out of the ensemble. This, however, requires a minimum fidelity $F_{\rm min}$ of the initial pairs to operate with, which cannot be achieved as the length $l$ of the channel increases.

In this paper, we treat the problem of a quantum repeater based on the method of entanglement purification. The idea of such a repeater is to divide 
a long quantum channel into shorter segments, which are purified separately, before they
are connected. Connecting two segments of a channel means here to build up quantum correlations across the compound channel from correlations that exist across the individual segments. This can be done by entanglement swapping, or teleportation of entanglement.
A quantum repeater must therefore combine the methods of entanglement purfication and teleportation. 

Although the combination of these methods 
should, in principle, allow to create entanglement over arbitrary distances, it is 
another question how much this "costs" in terms of resources needed for purification.
Resources means here the number of low fidelity EPR pairs that have to be provided 
for purification of each channel segment.
This quantity is related to the number of particles that have to be 
manipulated locally (at the connection points between the segments) in a coherent fashion.
If the resources grow too fast with the length of the channel, not much will be gained by the whole procedure.  
A further important quantity is the error tolerance for the local operations. In every real situation, the local operations applied to one or more particles, will bear some imperfections. Since such operations are the building blocks for any entanglement purification protocol, their imperfections will limit the maximum attainable fidelity for an EPR pair and the efficiency of the protocol. In the context of the quantum repeater, a
maximum fidelity $F<1$ corresponds to a residual amount of noise for each segment. When 
the segments are connected, this noise accumulates. 

To give a realistic treatment of the quantum repeater, we will therefore first study the applicability and the efficiency of entanglement purification protocols in the situation of imperfect local operations.
The general conditions under which standard purification protocols can be used in the presence of errors have been studied by Giedke {\em et al.} \cite{giedke98}. This includes, in particular, thresholds and lower bounds for the attainable fidelities. 
For a generic class of stochastic errors, one can give explicit representations of the imperfect operations in terms of completely positive maps and/or POVMs \cite{briegel98}.
In terms of these maps, we derive recursion formulas for the fidelities, which generalize the results given by Bennett {\em et al.} \cite{bennett96a} and by Deutsch {\em et al.} \cite{deutsch96}. From these results, we estimate that, for a generic class of errors, standard entanglement purifcation protocols work even for error
probabilities of the order of a few per cent. In the context of long distance communication, we develop 
a purification procedure that combines the standard protocols with quantum teleportation 
in a certain way to be specified. This purification procedure allows quantum communication via noisy channels of arbitrary length. Since it explicitly exploits two-way classical communication \cite{bennett96b}, our procedure is much more efficient for quantum communication than protocols based on quantum error correction \cite{knill96}.
Specifically, this solution of the quantum repeater tolerates errors on the per-cent level; 
it requires a polynomial overhead in time and an overhead in local resources that grows only logarithmically with the length of the channel.

The paper is organized as follows: In Sec.~\ref{Error} we introduce a generic error
model to describe imperfect local operations such as noisy 1 bit and 2 bit operations 
and imperfect measurements. With the aid of these models, we revisit in Sec.~\ref{SECelements} the basic elements of quantum communication --
purfication and teleportation. This discussion includes results for the maximum attainable fidelities and the working conditions of the standard protocols under the condition of noisy local operations. After this discussion, in Sec.~\ref{SECqrep}, we are ready to attack the quantum repeater problem itself. Our solution combines the methods of entanglement purification and entanglement swapping into a single (meta-) protocol, which we call nested entanglement purification. This protocol allows to distribute entanglement with a given
fidelity across arbitrary long distances, even when local operations are imperfect.
It requires an overhead in local resources that grows polynomially with the length
of the channel. A modification of this protocol works with resources that grow only logarithmically with the length of the channel (while the build-up time grows  polynomially). 


\section{Imperfect local operations and measurements}\label{Error}

Imperfections of local operations can have various origins. An example for an imperfect 
one-qubit operation is a unitary `overrotation' \cite{zurek96}, as it would be realized 
by a laser pulse that effects a Bloch rotation around an angle that slightly deviates from the ideal (intended) one. If the laser is well stabilized but the pulse area tuned incorrectly, say, the deviation of the real from the intended (unitary) rotation will be the same when we repeat the operation, representing an example of a systematic error.  If, on the other hand, the angle of the rotations fluctuates randomly around a certain mean value, every pulse introduces a certain amount of noise into the system, and
the resulting operation would be described by a non-unitary map. In the present paper, we concentrate on this latter situation, that is we assume only stochastic errors and neglect systematic errors.

To give a precise description of a noisy 1-bit or 2-bit operation, one needs to know the exact form of the error mechanisms, which in turn depend on the specific physical implementation. In general, we have only limited knowledge of these details. 
A generic model for a noisy channel is the so-called (symmetric) depolarizing channel. 
It transforms a qubit with initial state $\rho$ in the way
\begin{equation}
 \rho \longrightarrow p\,\rho + \frac{1-p}{2}I\,,
\label{depol}
\end{equation}
where $I$ is the identity operator and $0\le p\le 1$ depends on the time for which the qubit is subjected to the channel.  The action of the channel thus results in admixing a completely depolarized state $I/2$ to the initial density operator. The limit $p \to 0$ corresponds to a very noisy channel, while 
$p\to 1$ describes a channel with very little noise (``ideal storage").

One can easily generalize (\ref{depol}) to the case where the ideal operation 
is some arbitrary unitary (1-bit) operation $U$, 
\begin{equation}
 \rho \longrightarrow p\,U\rho\, U^{\dag}+ \frac{1-p}{2}I\,.
\label{gen_depol}
\end{equation}
In this situation, the ideal operation is accompanied by the action of the
depolarizing channel. The whole map becomes non-unitary. It describes an imperfect 1-bit operation, which for $p\to 1$ becomes identical to the ideal 
unitary map. We call $p$ the {\em reliability} of the imperfect operation, which represents a lower bound to the probability that its result corresponds to the ideal one.
For a parametrization $U=U_{\vec\alpha}=\exp(-i\vec\alpha\cdot\vec\sigma)$ an imperfect (noisy) rotation may explicitly be defined by the map
\begin{equation}
 \rho \longrightarrow R_{\vec\alpha}\rho = \int\text{d}\vec\beta\, \nu(\vec\beta) U_{\vec\alpha+\vec\beta}\rho\, U_{\vec\alpha+\vec\beta}^{\dag}\,.
\label{bloch_av}
\end{equation}
where the noise function $\nu(\vec\beta)$ is normalized to unity and describes
the fluctuations $\delta\vec\alpha = \vec\beta$ of the (generalized) Bloch vector 
$\alpha$. One can easily show that for ``isotropic noise," $\nu(\vec\beta) = \tilde\nu(|\vec\beta|)$, the integral reduces to 
\begin{eqnarray}
 R_{\vec\alpha}\rho &=& p\, U_{\vec\alpha}\rho\, U_{\vec\alpha}^{\dag} + \frac{1-p}{2}I \nonumber\\
 &=& p\, R_{\vec\alpha}^{\text{ideal}} \rho  + \frac{1-p}{2}I
\label{result_bloch_av}
\end{eqnarray} 
wherein $R_{\vec\alpha}^{\text id}\rho \equiv U_{\vec\alpha}\rho U_{\vec\alpha}^{\dag}$ describes the ideal rotation.

Motivated by these observations, we generalize this model to arbitrary 
one-qubit and two-qubit gates \cite{footnote1} acting on an entangled state $\rho $ of several qubits. 
We thereby require that errors introduced by local operations may only affect the local qubits on which the ideal operation acts (locality condition). Let $\rho $ be a state of $q$ qubits, labeled by $1,2,\dots,q$. An imperfect one-qubit operation acting on the first qubit, say, is then described by the map
\begin{equation}
 O_1 \rho = p_1 O_1^{\text{ideal}}\rho 
 + \frac{1-p_1}{2}\text{tr}_1\{\rho \} \otimes  I_1
\label{one_qubit_op}
\end{equation}  
whereas an imperfect two-qubit operation acting on qubits 1 and 2 is described by 
\begin{equation}
 O_{12} \rho = p_2 O_{12}^{\text{ideal}}\rho 
 + \frac{1-p_2}{4}\text{tr}_{12}\{\rho\} \otimes  I_{12}\,.
\label{two_qubit_op}
\end{equation}  
In these expressions, $O^{\text{ideal}}$ is the ideal (perfect) operation and
$I_1$ and $I_{12}$ denote unit operators on the subspace where the ideal operation acts, 
corresponding to totally depolarized one and two-qubit states. The state of the 
other particles are described by the partial traces $\text{tr}_1\{\rho \}$ and
$\text{tr}_{12}\{\rho \}$, respectively, of the initial density operator over theses subspaces. The quantities $p_1$ and $p_2$ measure the {\em reliability} of the operations, where perfect operations correspond to $p_j=1$. Technically speaking, an imperfect operation is modeled by the corresponding ideal operation accompanied/followed by a depolarizing channel that acts (only) on the same subspace as the ideal operation. 

Note that the maps (\ref{one_qubit_op}) and (\ref{two_qubit_op}) are linear and trace conserving, albeit non-unitary. For any state $\rho$ which is diagonal in the Bell basis defined by any two of the particles -- all states we deal with in this paper are of this form --, the model is self consistent in the sense that $O_1\otimes O_2$ (two single-qubit operations, each described by an error parameter $p_1$) can be written as a certain two-qubit operation $O_{12}$ described by an error parameter $p_2=p_1^2$. Generally, any sequence of local operations can be written as a single joint operation within this model. The resulting error parameter (i.e.\ reliability) describing the joint operation is obtained by multiplying the error parameters of the single operations. 

We will use these maps to estimate the role of imperfect operations in 
standard protocols for entanglement purification and for teleportation.
Although the model is simple, it is generic at the same time. We just mention here that 
similar maps may be derived from a Jaynes-type principle, insofar as their resulting states maximize the entropy of the resulting state with respect to variations that are trace conserving and keep the average fidelity of the operation constant. 

We finally describe an {\em imperfect measurement} on a single qubit (in the computational basis) by a POVM \cite{helstrom76} corresponding to the `imperfect projectors'
\begin{eqnarray}
 P_0^{\eta} &=& \eta |0\rangle\langle 0| + (1-\eta ) |1 \rangle\langle 1|\nonumber \\
 P_1^{\eta} &=& \eta |1\rangle\langle 1| + (1-\eta ) |0 \rangle\langle 0|\,.
\label{povm}
\end{eqnarray}
The parameter $\eta $ is a measure for the quality of the projection onto
the basis states. Assume the qubit to be measured is in the state 
$\rho =|0\rangle\langle 0|$ and we are trying to measure its state with the aid of an
measurement apparatus described by (\ref{povm}). The expectation values $\langle P_0^{\eta} \rangle = \eta$ and $\langle P_1^{\eta}\rangle = 1-\eta$ simply mean that the apparatus will give us the wrong result (``1'') with probability $1-\eta \ge 0$. That is, the result is not
completely reliable. An ideal measurement, in contrast, is described by $\eta=1$. It is clear that the effect of the measurement on the measured qubit is not fully specified 
by the POVM (\ref{povm}). In the present context, however, this description is sufficient
as all measured particles are removed from the system i.e.\ we trace over their degrees of freedom after the measurement. 

The operations (\ref{one_qubit_op}) and (\ref{two_qubit_op}) together with single-qubit measurements (\ref{povm}) are sufficient to describe all operations occurring in the context of teleportation and entanglement purification. For example, a Bell measurement (the measurement of a projector in the Bell basis) on two particles, say 1 and 2, can be realized by a two-qubit operation $O_{12}\equiv \mbox{CNOT}_{1\to 2}^{\text{imperf}}$ (controlled NOT) followed by a Hadamard rotation of particle 1 and two single-qubit measurements $P_{0(1)}^{1}$, $P_{0(1)}^{2}$ on particles 1 and 2 \cite{footnote3}. Instead of performing the Hadamard transformation one can also measure particle 1 in a rotated basis. In summary, an imperfect Bell measurement is described by an imperfect two-qubit operation followed by 2 imperfect single-qubit measurements, effecting
e.g.\ $|0\rangle|0\rangle\pm|1\rangle|1\rangle \rightarrow (|0\rangle\pm|1\rangle)|0\rangle$.


\section{Purification and teleportation with imperfect means}\label{SECelements}

In this section we reconsider the basic elements of quantum communication, which are
teleportation and purification, in the presence of local errors.

\subsection{Entanglement purification}\label{Purification}

Purification is the distillation of few 'perfect' EPR pairs out of many imperfect pairs. In the following, we will generalize two different purification protocols that have been treated in the literature, by introducing imperfect gate and measurement operations. 
In subsection \ref{PurificationC}, we then discuss a modified version of these schemes which, generally speaking, has less favorable convergence properties but works with a smaller and constant number of {\it physical resources}.

\subsubsection{Scheme of Bennett {\em et al.} (Scheme A)}

\begin{figure}[ht]
\begin{picture}(150,67)
\put(0,0){\epsfysize=59pt\epsffile[0 0 140 57]{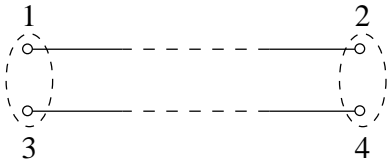}}
\end{picture}
\caption[]{Particles required for purification in standard recurrence schemes.}
\label{purify1}
\end{figure}

This purification scheme was introduced by Bennett {\em et al.} \cite{bennett96a,bennett96b}. 
In short, the scheme takes two pairs (1-2 and 3-4) as in Fig.~\ref{purify1}, both in a Werner state 
\be 
\rho=F|\Phi^+\rangle\langle\Phi^+|+\left(\frac{1-F}{3}\right)\left(|\psi^{-}\rangle\langle\psi^{-}|+|\psi^+\rangle\langle\psi^+|+|\Phi^-\rangle\langle\Phi^-|\right)\,. 
\label{Werner} \ee
with fidelity $F=\langle\phi^+| \rho |\phi^+\rangle$.
Then it performs local (1 $\&$ 2-bit) operations on the particles at the same ends of the pairs.
Writing $\rho_{12}\otimes\rho_{34}=\rho_{1234}$, this essentially 
involves two CNOT operations, CNOT$_{13}^{\text{imperf}}$ and CNOT$_{34}^{\text{imperf}}$ 
acting on the state $\rho_{1234}$, followed by a simple measurement 
$P_{0(1)}^{3}$, $P_{0(1)}^{4}$  of the particles $3$ and $4$. If the particles are found in 
the same state (00 or 11), the remaining pair, described by the state  $\rho_{12}'$ is kept, otherwise it is discarded. To obtain a recursive mapping between Werner states, a non-unitary depolarization operation (twirl) is applied to the resulting state before it is
used for the next purification step. The original treatment of this protocol \cite{bennett96a,bennett96b} assumes that all operations and measurements of the protocol
are perfect.

To estimate the effect of local errors, we evaluate the protocol with the aid of the
imperfect operations (\ref{one_qubit_op}--\ref{povm}). 
The calculation of the fidelity $F'$ of the new pair  $\rho_{12}'$ is somewhat lengthy but straightforward, and the result can be given in analytic form. The result of
this calculation is summarized in the formula  
\widetext
\begin{equation}
 F' = \frac{[F^2+(\frac{1-F}{3})^2][\eta^2+(1-\eta)^2]
       +[F(\frac{1-F}{3})+(\frac{1-F}{3})^2)][2\eta(1-\eta)]
       +(\frac{1-p_2^2}{8p_2^2})}{[F^2+\frac{2}{3}F(1-F)
       +\frac{5}{9}(1-F)^2][\eta^2+(1-\eta)^2]
       +[F(\frac{1-F}{3})+(\frac{1-F}{3})^2)][8\eta(1-\eta)]
       +4(\frac{1-p_2^2}{8p_2^2})}
\label{modified_bennett}
\end{equation}
which reduces to the formula given in Ref.~\cite{bennett96a} in the limiting case $\eta=1$ and $p_2=1$ (perfect operations). Although $\rho_{12}'$ is no longer a Werner state, it can again be brought to Werner form using depolarization, which is considered to be noiseless here \cite{footnote2}.
If one starts with an ensemble of pairs with fidelity $F$, $F'$ gives the fidelity of the
remaining pairs that are left after one purification step. This defines a fraction 
of $2/p_{\text{even}}$ of the initial ensemble of pairs, where $p_{\text{even}}$ denotes the 
probability for finding the particles $3$ and $4$ in the state 00 or 11, and is given 
by the denominator in (\ref{modified_bennett}). The factor 
2 arises since the pair originally composed of particles $3$ and $4$ is lost due to 
the measurement. 

Each (iterated) application of (\ref{modified_bennett}) corresponds to a purification step. For each purification step, 2 identically pairs (which both result from previous successful purification steps) are used. The {\em purification resources} are defined by the  average number $M$ of pairs needed to perform $k_{max}$ successful purification steps, and are given by
\begin{equation}
 M = \prod_k^{k_{\text{max}}} \frac{2}{p^{(k)}_{\text{even}}}\,,
\label{purisources}
\end{equation} 
where the probabilities $p^{(k)}_{\text{even}}$ depend on the purification step $k$.
The {\it physical resources}, defined by the total number of particles at A or B that are
used to store the pairs needed for the purification process, are for this scheme equal to the purification resources, since all pairs are created at the beginning and have to be stored as can be seen in Fig.\ref{tree}.

\begin{figure}[ht]
\begin{picture}(230,220)
\put(0,0){\epsfxsize=200pt\epsfysize=180pt\epsffile[17 45 523 684 ]{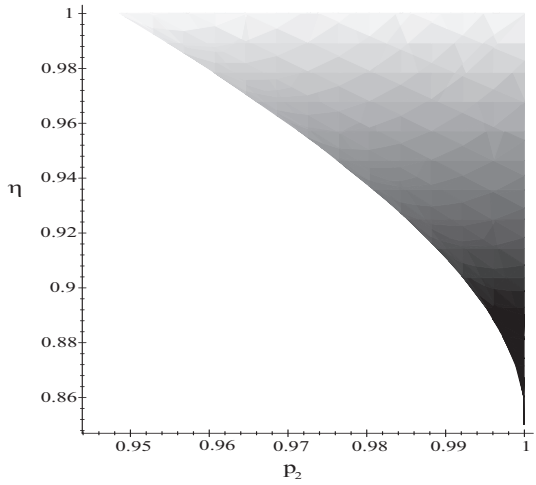}}
\end{picture}
\caption[]{Region for error parameters $p_2$ and $\eta$ where purification is possible (scheme A)}
\label{FixB3}
\end{figure}

\begin{figure}[ht]
\begin{picture}(230,200)
\put(0,0){\epsfxsize=230pt\epsffile[35 175 561 605 ]{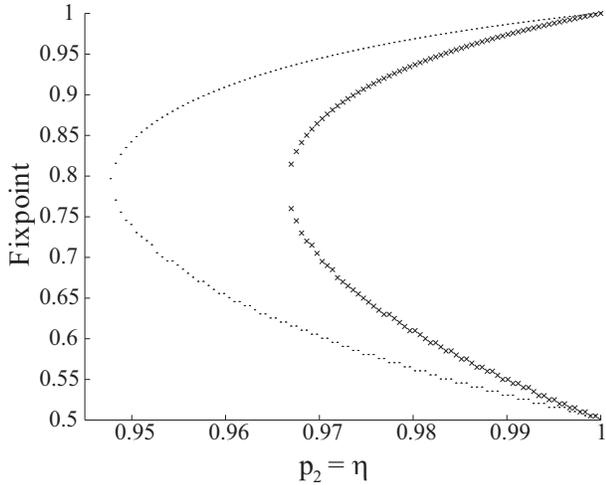}}
\end{picture}
\caption[]{Fixpoint ($F_{\rm min}$ and $F_{\rm max}$) of scheme A ($\times$) and scheme B ($\bullet$) plotted against error parameters $p_2=\eta$.}
\label{BenOxf3}
\end{figure}

When calculating the fixpoints of (\ref{modified_bennett}), one finds that one fixpoint is always $F=\frac{1}{4}$, independent of the error parameters $p_2$ and $\eta$. The other two fixpoints $F_{\rm min}$ and $F_{\rm max}$ (see Fig.~\ref{FIGpuriloop}) are given by the 
expression
\begin{equation}
 F_{\text{max} \atop \text{min}} = \frac{8\eta(\eta-1)+3\pm \sqrt{10-9/p_2^2
 +64\eta^4-128\eta^3+116\eta^2-52\eta-36\eta(\eta-1)/p_2^2}}{16\eta(\eta -1)+4}\,.
\label{fixpoints}
\end{equation}
They depend on the error parameters and give the borders of the interval within which purification is possible. If $F\in(F_{\rm min},F_{\rm max})$, then $F'> F$. Since $F_{\rm max}$ is an attractor, iterative application of (\ref{modified_bennett}) leads to a resulting pair with fidelity $F\rightarrow F_{\rm max}$. The value $F_{\rm min}$ thus gives the threshold for F where this purification scheme can be successfully applied, while $F_{\rm max}$ gives the maximal reachable fidelity. For perfect local operations, $F_{\rm min}=\frac{1}{2}$ and $F_{\rm max}=1$, meaning that all pairs with $F>\frac{1}{2}$ can be purified to $F=F_{\rm max}=1$. For imperfect local operations, $F_{\rm min}>\frac{1}{2}$ and $F_{\rm max}<1$, i.e. no perfect EPR pairs can be created. If the error parameters become larger, these two fixpoints approach each other and the region where purification is possible shrinks to zero. The limiting situation $F_{\rm max} = F_{\rm min}$ 
defines the threshold for the applicability of the purification protocol.
For all pairs ($p_2,\eta$) for which there is only one real fixpoint 
(at $F=1/4$), the imperfections of the local operations introduce more noise 
than one gains from the purification, so the scheme breaks down.
For example for $\eta=1$, (perfect measurements), Eq.~(\ref{fixpoints}) simplifies to 
\begin{equation}
 F_{\text{max} \atop \text{min}} = \frac{3}{4}\pm\frac{1}{4}\sqrt{10-9/p_2^2}
\label{simple_fixpoints}
\end{equation}
with the threshold at $p_2 = \sqrt{9/10}\simeq 0.95$. That is, the CNOT gate must work with a reliability of at least $95\%$.  The threshold gets tighter when the measurements are imperfect as well, i.e.\ for $\eta < 1$. In general, the threshold values for the error parameters $p_2$ and $\eta$ are of the order of some percent as can be seen in Fig.\ref{FixB3}. The fixpoints as a function of the error parameters are plotted in Fig.\ref{BenOxf3}, where we set $p_2=\eta$ to get a 2 dimensional plot.

\subsubsection{Scheme of Deutsch {\em et al.} (Scheme B)}

A purification protocol that converges faster and involves less resources was proposed by 
Deutsch {\em et al.} \cite{deutsch96}. Generally speaking, this scheme can be described as a mapping on states that are diagonal in the Bell basis, but need not necessarily be Werner states,
\begin{equation}
 \rho_{12}=A\,|\phi^{+}\rangle\langle\phi^{+}| + B\,|\psi^{-}\rangle\langle\psi^{-}|+  
 C\,|\psi^{+}\rangle\langle\psi^{+}| + D\,|\phi^{-}\rangle\langle\phi^{-}| 
\label{bell_diag}
\end{equation} 
where $|\phi^{\pm}\rangle$ and $\psi^{\pm}\rangle$ are the Bell states between particles 1 and 2 in the usual notation. In the notation of (\ref{bell_diag}), the protocol corresponds to a mapping $R^4\rightarrow R^4$ between the diagonal elements (A,B,C,D).
In particular, between two successive purification steps, the states are not depolarized. This feature, together with additional $\pi/2$ rotations that are applied to the qubits before the CNOT operations, makes scheme B a much faster converging protocol.        

In the following analysis, we evaluate scheme B using the imperfect operations (\ref{one_qubit_op})--(\ref{povm}) instead of perfect ones.  
A $\pi/2$ rotation followed by a 
CNOT operation is thereby treated as a joint 2-qubit operation with a single error parameter $p_2$. The resulting map between the diagonal elements $R^4\rightarrow R^4$ is described in \cite{Duer98}; it completely characterizes the action of this purification protocol.
Successive purification steps are now described by iterated applications of this map. For each purification step, two identical pairs (which result from the previous step) are used. The resources can be calculated in a similar way as for scheme A. 

The fixpoints of this map are no longer described by a single parameter F but by a set of 4 numbers $(A_{fix},B_{fix},C_{fix},D_{fix})$. The fidelity of such a diagonal Bell state is given by $A$, the $|\phi^+\rangle$ component and, for simplicity, we shall continue to call this component ``fixpoint" of scheme B, although the fixpoint is not sufficiently described by this single parameter. For example, it may happen that a Werner state with a certain fidelity $F_0$ cannot be purified, while a binary state with the same fidelity $F_0$ that
has only two nonzero components $|\phi^+\rangle$ and $|\psi^+\rangle$, say, (in short: "binary state (A,C)") can be purified. 

\begin{figure}[ht]
\begin{picture}(230,200)
\put(0,0){\epsfxsize=230pt\epsffile[18 177 552 593]{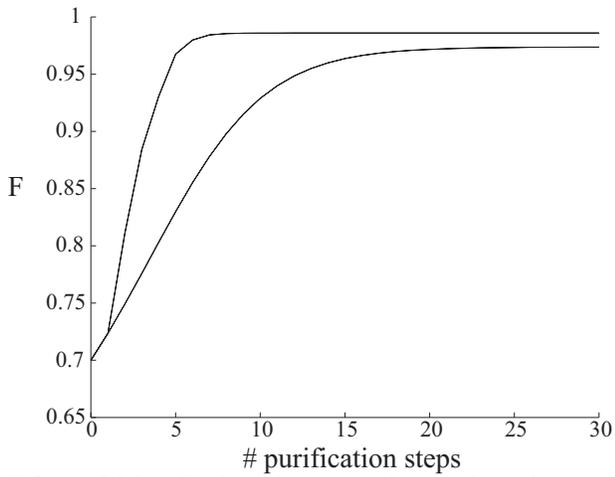}}
\end{picture}
\caption[]{fidelity F plotted against the number of successful purification steps for scheme B (upper curve) and scheme A. Based on fixed error parameters $p_2=\eta$ of 1\% and initial fidelity $F_0=0.7$.}
\label{BenOxf1}
\end{figure}

We have numerically compared schemes A and B, and found that scheme B converges much faster towards the upper fixpoint $F_{\rm max}$ if $F_0 > F_{\rm min}$ as can be seen in Fig.\ref{BenOxf1}. Surprisingly, the upper fixpoint for scheme B is above the one for scheme A for given error parameters and does not depend on the ``shape" of the initial state. The lower fixpoint $F_{\rm min}$ is smaller for scheme B and therefore the interval within which purification is possible is {\it larger} for scheme B, and so is the maximal reachable fidelity. The fixpoints as a function of the error parameters $p_2=\eta$ are plotted in Fig.\ref{BenOxf3}. One can see that scheme B is significantly less sensitive to noisy 
local operations than scheme A.

\subsubsection{A modified purification scheme (scheme C)}\label{PurificationC}

Both schemes (A and B) need many elementary pairs to start with, which have to be stored somewhere before the protocols are applied. The physical resources are thus quite large and grow with the number of necessary purification steps. In the following, we modify 
scheme B in a certain way, and call this scheme C.

\begin{figure}[ht]
\begin{picture}(230,30)
\put(0,0){\epsfxsize=230pt\epsffile[0 0 349 49]{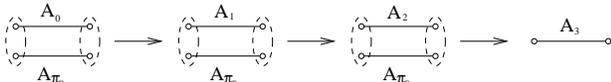}}
\end{picture}
\caption[]{Typical purification process. At each step, the same elementary pair $\pi_0$ is used. If one purification step is not successful, one has to start from the beginning.}
\label{IbkResfig}
\end{figure}

Different from scheme B, we do not use two identical pairs at each purification step (which are left over from previous purification steps) but always purify one and the same pair 
with the help of an auxiliary pair $\pi_0$ with constant fidelity $A_{\pi_0}$. Apart from that, we employ the same protocol (that is the sequence of local operations and measurements) as in scheme B. The resulting map is thus the same, the only difference being that $\rho_{34}$ is a diagonal Bell state which is constant throughout the whole purification procedure and is given by $(A_{\pi_0},B_{\pi_0},C_{\pi_0},D_{\pi_0})$. This auxiliary pair is repeatedly created before each purification step between two of the four particles as visualized in Fig.\ref{IbkResfig}. If a purification process is not successful, one has to start from the beginning (when both pairs 12 and 34 are put in the state $\pi_0$).
Using this procedure, the fidelity converges towards a given fixpoint $F(\pi_0)$ which is {\it not} equal to one \cite{Luetkenhaus}. The fixpoint highly depends on both the initial fidelity {\it and} the shape (other diagonal elements in the Bell basis) of the state $\pi_0$. The properties of the fixpoint will be discussed in more detail later on.

\begin{figure}[ht]
\begin{picture}(230,200)
\put(0,0){\epsfxsize=230pt\epsffile[19 166 548 593 ]{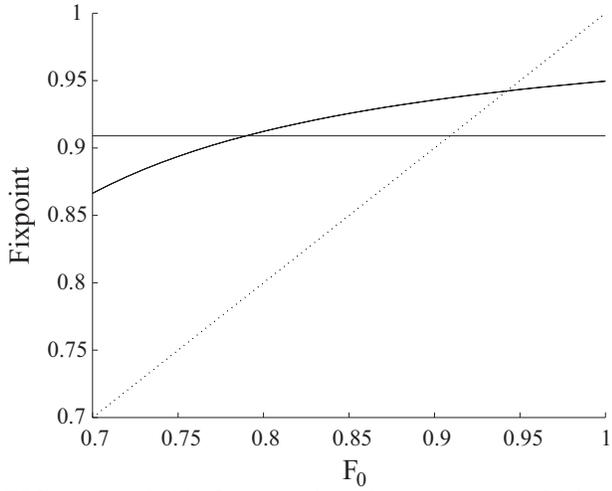}}
\end{picture}
\caption[]{Reachable fixpoint plotted against initial fidelity $F_0$  for scheme B (constant function) and scheme C. Based on fixed error parameters $p_2=\eta$ of 4\% and binary pairs (A,C).}
\label{IbkOxf1}
\end{figure}

For ideal operations, this method is generally less favorable as far as its convergence properties are concerned. However, for imperfect operations, the situation changes and the drawback that the reachable fixpoint $F_{\pi_0}$ is smaller than unity becomes less important. As we shall see later, for imperfect local operations the fixpoint of scheme C may lie even {\em above} the fixpoints of schemes A and B (see Fig.~\ref{IbkOxf1}).
The main advantage of this scheme is, however, that the {\it physical resources}, that is the number of particles needed at A or B to store the pairs, are {\it constant} (namely 2) and independent of the number of necessary purification steps. 
This fact makes scheme C particulary interesting for the repeater problem, where the accumulation of local particles at the connection points of a compound long channel plays an important role. 

The {\em purification resources}, on the other hand, that is the (average) number of how often an auxiliary pair has to be created in order to perform $k_{max}$ purification steps, 
have now to be calculated in a different way.
Let $p_{even}(k)$ be the probability to succeed at the $k^{th}$ purification step. The total resources can be calculated using the iteration formula
\be M_{k+1}=(M_k+1)\frac{1}{p_{even}(k+1)}\,. \ee
To close this formula, note that $M_0$=1. This can be understood as follows: Starting at the elementary level, one needs two pairs for purification and has to repeat the procedure on average $\frac{1}{p_{even}(1)}$ times. Thus $M_1=(1+1)\frac{1}{p_{even}(1)}$. For the next purification step, one needs one additional elementary pair to perform the next purification step. This step has to be performed on average $\frac{1}{p_{even}(2)}$ times. The resources to perform the second purification step are thus given by $M_2=(M_1+1)\frac{1}{p_{even}(2)}$ and so on. To perform $k_{max}$ purification steps, one needs to create
on average 
\be M_{k_{max}}=\sum_{i=1}^{k_{max}}\left(\prod_{k=i}^{k_{max}}\left(\frac{1+\delta_{k1}}{p_{even}(k)}\right) \right) 
\label{iteration}
\ee     
times an auxiliary pair.

A related quantity is the average number of purification steps (including also those which fail). This is given by the iteration formula
\be S_{k+1}=(S_k+1)\frac{1}{p_{even}(k+1)} \ee
where $S_0=0$ and k denotes the number of {\it successful} purification steps. If no purification is performed, the number of steps $S_0=0$. This is the only difference to the iteration formula (\ref{iteration}) for the resources, since every time an elementary pair is created, one performs a purification step. Thus the average total number of purification steps performed (assuming that $k_{max}$ steps were successful) is given by
\be S_{k_{max}}=\sum_{i=1}^{k_{max}}\left(\prod_{k=i}^{k_{max}}\left(\frac{1}{p_{even}(k)}\right) \right)\,.\label{iteration1}\ee 

\begin{figure}[ht]
\begin{picture}(200,200)
\put(0,0){\epsfxsize=230pt\epsffile[25 182 548 606 ]{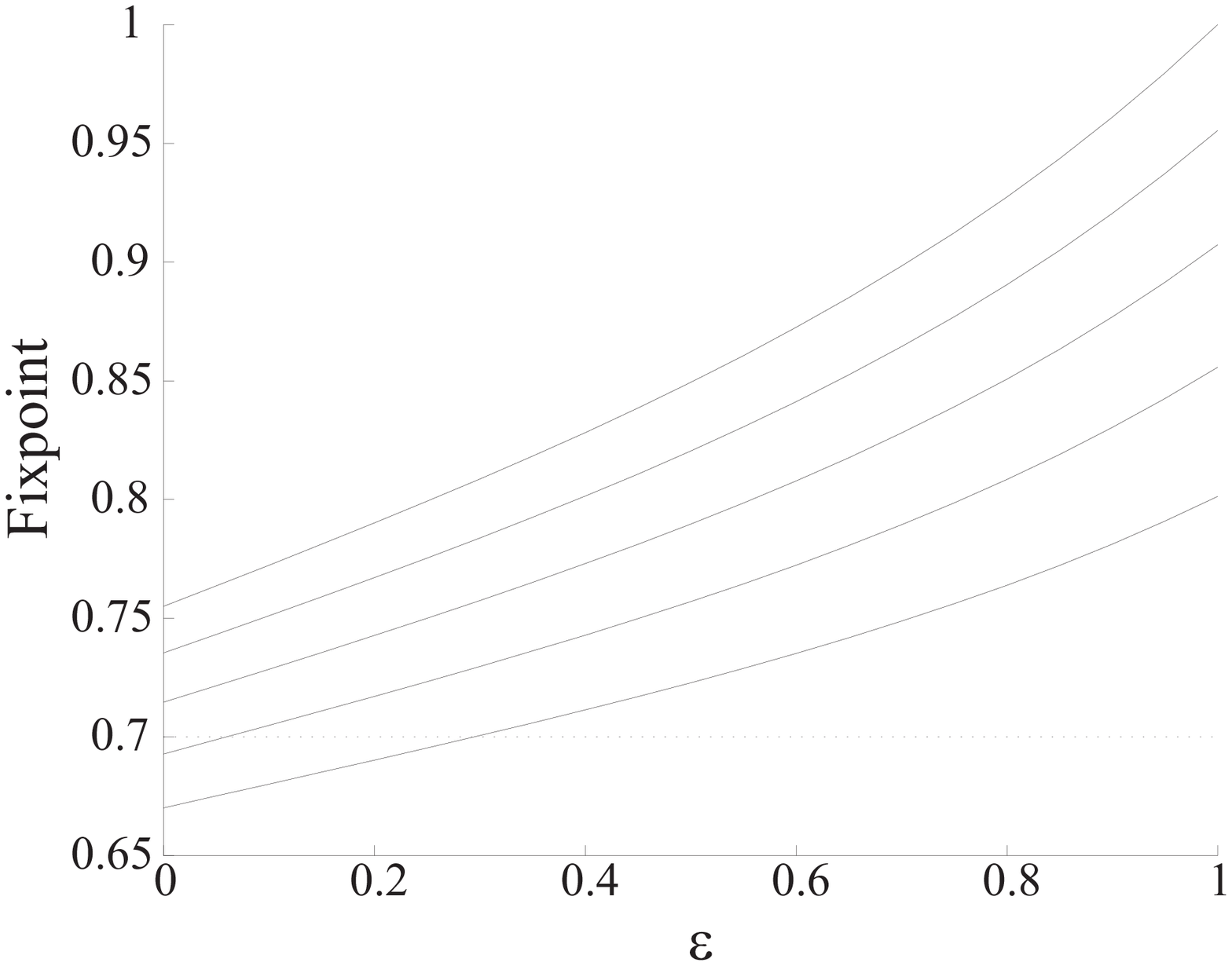}}
\end{picture}
\caption[]{Reachable fixpoint plotted against parameterization parameter $\epsilon$ (shape of the state) for fixed initial fidelity $F_0=0.7$. Plots from bottom to top correspond to error parameters $p_2=\eta$ of 4\%, 3\%, 2\%, 1\%, 0\%.}
\label{Ibk1}
\end{figure}

The fixpoint discussion for this purification protocol is more involved due to the many parameters, namely the fidelity $A_{\pi_0}$ and the shape (other diagonal elements $B_{\pi_0},C_{\pi_0},D_{\pi_0})$) of the elementary pair, and the error parameters $p_2$ and $\eta$ describing the noisy local operations. For this protocol, purification is possible if the reachable fidelity (fixpoint) for a certain auxiliary pair lies above its initial fidelity. To illustrate first the shape dependence of the fixpoint, we introduce the following parameterization of diagonal Bell states with constant fidelity $F_0$. In the notation 
of (\ref{bell_diag}) we write
\bea
A&=&F_0 \nonumber \\
B=C&=&(1-F_0)(1-\epsilon)/2 \nonumber \\
D&=&(1-F_0)\epsilon 
\eea
where $0\leq \epsilon \leq 1$. A Werner state thereby described by $\epsilon=\frac{1}{3}$ and a binary state (A,D) is obtained for $\epsilon=1$. Figure \ref{Ibk1} shows the dependence of the fixpoint as a function of $\epsilon$ for different error parameters.     

For the manifold of states covered by this parametrization one sees that the 'optimum shape' of the elementary pair is a binary state (A,D), meaning that the reachable fixpoint for a given initial fidelity is the largest for this shape. Figure \ref{IbkOxf1} demonstrates two different things. First, it shows that the fixpoint (maximally reachable fidelity) depends on the initial fidelity of the elementary pair even if the other components are kept constant.  
Second, it demonstrates that the reachable fidelity using this purification scheme may lie above the reachable fidelity using schemes A or B. (The upper fixpoint for scheme A is not plotted here, but as we already know (see Fig.\ref{BenOxf3}) it always lies below the fixpoint for scheme B).  
One should however mention that this is only the case if the shape of the elementary pair is close to the optimal configuration (A,D). This can be understood as follows: Scheme B always picks 2 pairs that result from a previous purification step and the shape of such a pair converges towards a ``working state" which no longer depends on the initial shape but only on the error parameters. Scheme C, on the other hand, always uses the same auxiliary pair for each purification step which has {\it not} been influenced by noisy local operations
in previous steps. As we have pointed out earlier, in general one can only increase the fidelity by a certain amount. 

\subsection{Teleportation}\label{Connection}

\begin{figure}[ht]
\begin{picture}(230,25)
\put(0,5){\epsfxsize=230pt\epsffile[0 0 257 19]{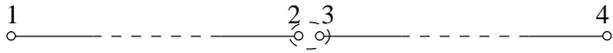}}
\end{picture}
\caption[]{Particles involved in the connection process. Particle 1 is located at A, particles 2 and 3 at $C_1$ and particle 4 at B.}
\label{teleport_2}
\end{figure}

In the context of the quantum repeater, teleportation comes into play when 
two (purified) segments of a channel are connected. The channel segments are 
represented by EPR pairs between particles 1-2 and 3-4 that have been created across the corresponding segments, see Fig.~\ref{teleport_2}. In general, to teleport the state of particle 2 to particle 4, a Bell measurement has to be made on particles 2 and 3 (and the result of this measurement 
communicated particle 4). If particle 2 is itself entangled to some other particle, say 
particle 1 as in Fig.~\ref{teleport_2}, the effect of the teleportation is to transfer
this entanglement to an entanglement between particles 1 and 4. This process has 
also been termed ``entanglement swapping" \cite{entanglement_swapping}. 

As a result, one obtains an EPR pair between particles 1 and 4 which can be used as a single
channel (again with the help of teleportation). In this sense, the segments of the compound 
channel have been connected. Put in different terms, the connection process creates quantum correlations across the compound channel from correlations that exist across the individual segments. 

We will study this process first for the particular case where the involved pairs (1-2 and 3-4) are not maximally entangled but are both in a Werner state (\ref{Werner}) with fidelity F.  
To connect the pairs (1-2) and (3-4), we make a Bell measurement on particles 2 and 3, which is realized by a (imperfect) CNOT operation followed by two (imperfect) single-particle measurements, as described in the last paragraph of Section \ref{Error}. 
Depending on the outcome of this measurement, a one-particle operation $O_4$ is performed on particle 4, as in the teleportation scheme \cite{bennett93}. The measurement outcome needs
thereby to be sent to particle 4 using classical communication. The result of this sequence of operations, followed by a subsequent depolarization, is a Werner state between particle 1 and 4 with a smaller fidelity $F_2$, which is independent of the outcome of the measurement. 

\begin{figure}
\begin{picture}(230,30)
\put(0,0){\epsfxsize=230pt\epsffile[0 0 751 66]{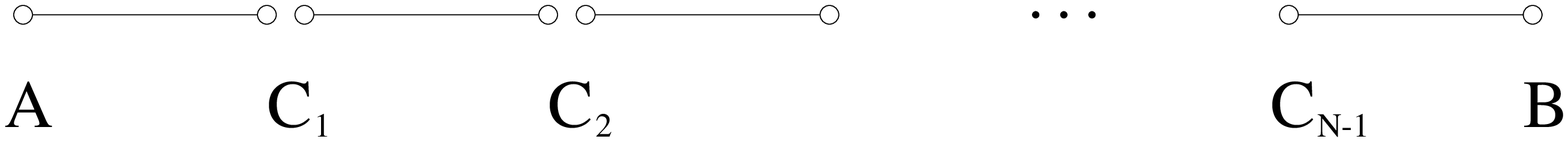}}
\end{picture}
\caption[]{Connection of a sequence of $N$ EPR pairs.}
\label{FIGmanypairs}
\end{figure}

More generally, imagine that we have not only 2 pairs but a whole string of $N$ pairs as visualized in Fig.~\ref{FIGmanypairs}. Each of the pairs ($A-C_1$),($C_1-C_2$),....($C_{N-1}-B$) is 
assumed to be in a Werner state with fidelity F. Connecting these $N$ pairs using the procedure described above leads to a pair ($A-B$) with fidelity $F_N$ given by the formula
\begin{equation}
 F_N = \frac{1}{4}\left\{1+3(p_1p_2)^{N-1}\left(\frac{4\eta^2-1}{3}\right)^{N-1} 
\left(\frac{4F-1}{3} \right)^N\right\}\,.  
\label{connect}
\end{equation}
The parameters $\eta,p_1$ and $p_2$ appearing in this formula quantify the amount of noise that is introduced by the connection processes. The connection therefore leads to an 
exponential decrease of the resulting fidelity, unless both the elementary 
pairs and all the operations involved have unity fidelity. 

The connection can be performed in two different ways, which both lead to the same resulting state with fidelity $F_N$ but involve a different temporal ordering of the operations. The first way is to connect the pairs {\it sequentially}, i.e. first connect at $C_1$, then at $C_2$, and so on, each time only connecting one additional pair. This involves a sequence of $N-1$ connection procedures.
The second way is to connect the pairs in {\it parallel}. To achieve this, first connect simultaneously the neighboring pairs at $C_1$, $C_3$, ... $C_{N-1}$. This leaves us with
longer pairs $(A-C_2)$, $(C_2-C_4)$,..., $(C_{N-2}-B)$. Then connect simultaneously these 
longer pairs at $C_2$,$C_6$,...,$C_{N-2}$, and so on, until we get a final pair between A and B. To have at each step pairs of equal length and fidelity, $N$ should be some power of 2, $N=2^n$,
although this is not an essential requirement. This method is much faster as it requires {\it fewer} iteration steps, namely $log_2N=n$ instead of $N-1$, although the number of local connections to be made is, of course, the same.

The elementary pairs that are connected need not be in a Werner state.
For example, for Bell diagonal states one can derive a map $R^4 \rightarrow R^4$ that relates the elements $(A,B,C,D)$ 
of the corresponding states before and after the connection. That is, Bell diagonal
states are mapped to Bell diagonal states under connection. All other remarks about
the connection procedure for a string of pairs apply here as well. The parallel way of connecting the pairs is preferable not only because can be done faster, but also because, 
after each connection step, one deals with an ensemble of {\em identical} pairs. 
Further details are given in \cite{Duer98}.


\section{Quantum communication over long distances}\label{SECqrep}

\subsection{Concept of the quantum repeater: Nested entanglement purification}\label{Repeater}

We have now all necessary tools available to introduce the concept of the quantum repeater. Our goal is to create an EPR pair of high fidelity between two distant locations. Since nonlocal entanglement between distant particles cannot be created using only local operations, this involves the usage of a quantum channel, which is noisy in general. The bottleneck for communication over large distances is the scaling of the error probability with the length of the channel. When using, for example, optical fibers and single photons as a quantum channel, both the absorption losses and the depolarization errors scale {\it exponentially} with the length of the channel. The state of the photon or the photon itself will therefore be destroyed with almost certainty if the channel is longer than a few half-lengths of the fiber. 

To overcome this limitation, we divide the long channel into $N$ smaller segments and create less distant EPR pairs across each segment as visualized in Fig.~\ref{FIGmanypairs}. The number of segments $N$ is thereby chosen in such a way that it is possible to create EPR pairs with sufficiently high initial fidelity $F>F_{\text min}$ over the distance of such a segment. In a next step, we connect these ``elementary" pairs as described in Section \ref{Connection}. This leaves us with a pair between A and B with reduced fidelity $F_N$
as given in (\ref{connect}). In principle, one could now create many pairs between A and B in a similar way and then use this ensemble of pairs for purification. But purification is only possible if the fidelity of the initial pairs is above a certain threshold value $F_{\rm min}$ as we have seen in Section \ref{Purification}. This limits the number of of pairs one can connect before purification becomes impossible. We therefore connect a smaller number $L \ll N$ of pairs so that the resulting fidelity $F_L$ stays above the threshold value for purification ($F_L \geq F_{\rm min}$) and purification is possible. 

\begin{figure}
\begin{picture}(230,140)
\put(0,0){\epsfxsize=230pt\epsffile[0 0 550 319]{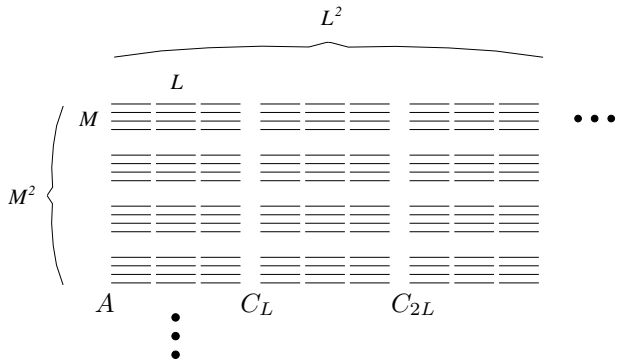}}
\put(33,18){$A$}\put(88,18){$C_L$}\put(145,18){$C_{2L}$}
\end{picture}
\caption[]{Nested purification with an array of elementary EPR pairs.}
\label{FIGnesting}
\end{figure}

The general strategy will be to design an alternating sequence connection and 
(re-)purification procedures in such a way that the number of resources needed 
remains as small as possible, and in particular does not grow exponentially with 
$N$ and thus with $l$. In the remainder of this Section we describe a {\em nested purification protocol} which consists of connecting and purifying certain groups of 
pairs simultaneously in the following
sense (see Fig.~\ref{FIGnesting}). For simplicity, assume that 
$N=L^n$ for some integer $n$. On the first level, we
simultaneously connect the pairs (initial fidelity $F_1$) at all the 
checkpoints except at $C_L,C_{2L},\ldots,C_{N-L}$. As a result, we have 
$N/L$ pairs of length $L$ (and fidelity $F_L$) between 
$A$ \& $C_L$, $C_L$ \& $C_{2L}$ and so on.

To purify these pairs, we need a certain number $M$ of copies that we
construct in parallel fashion. [For keeping track of the resources, it is
convenient to arrange them in form of an array of elementary pairs as is 
done in  Fig.~\ref{FIGnesting} for $L=3$ and $M=4$]. 
We then use these copies on the segments $A$ \& $C_L$, $C_L$ \& $C_{2L}$
etc., to purify and obtain one pair of fidelity $\ge F_1$ on each segment.
This last condition determines the (average) number of copies $M$ that we
need, which will depend on the
initial fidelity, the degradation of the fidelity under connections, and
the efficiency of the purification protocol. 
The total number of elementary pairs we used up to this point is $LM$. 
On the second level, we connect $L$ of these larger pairs at every 
checkpoint $C_{kL}$ ($k=1,2\ldots$) except
at $C_{L^2},C_{2L^2},\ldots,C_{N-L^2}$. As a
result, we have $N/L^2$ pairs of length $L^2$ between $A$ \& $C_{L^2}$,
$C_{L^2}$ \& $C_{2L^2}$, and so on of fidelity $\ge F_L$. Again, we 
need $M$ parallel copies of these long pairs to re-purify up to a fidelity 
$\ge F_1$. The total number of elementary pairs involved up to this point
is $(LM)^2$. We iterate the procedure to higher and higher levels, until we
reach the $n$--th level. As a result, we have obtained a final pair between
$A$ \& $B$ of length $N$ and fidelity $\ge F_1$. 
In this way, the total number $R$ of elementary pairs will be $(LM)^n$ 
[where $M^n$ alone gives the number of required `parallel channels' in 
Fig.~\ref{FIGnesting}]. 
We can re-express this result in
the form
\begin{equation}
 R = N^{\log_{L}M+1}
\label{resources}
\end{equation}
which shows that the resources grow {\em polynomially} with the distance $N$.

\begin{figure}
\begin{picture}(190,200)
\put(-40,10){\epsfxsize=190pt\epsffile[0 0 565 731]{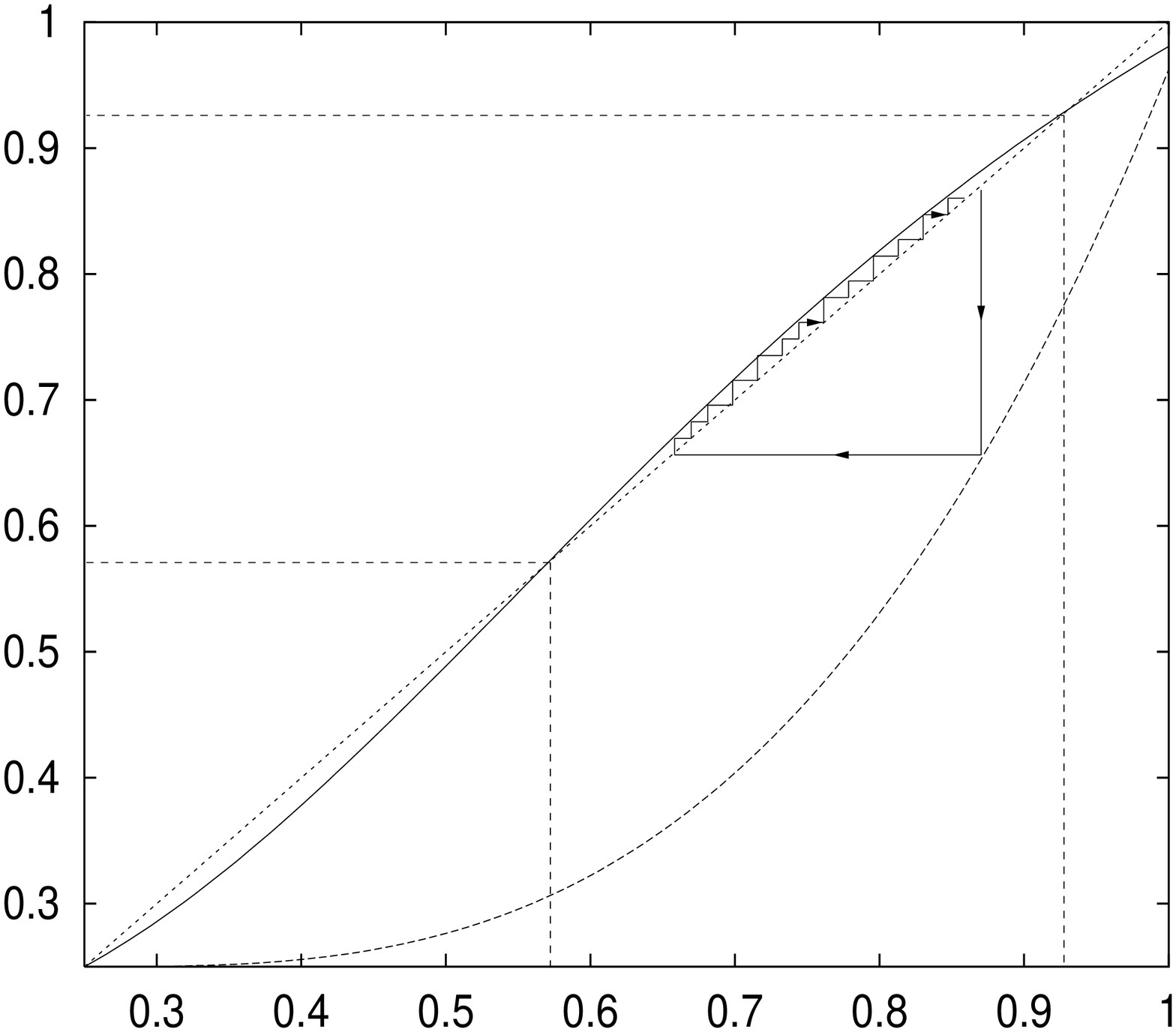}}
\put(97,0){$F$}
\put(120,150){$F'$}
\put(148,84){$F_L$}
\put(10,100){$F_{\text{min}}$}
\put(10,182){$F_{\text{max}}$}
\end{picture}
\caption[]{"Purification loop" for connecting and purifying EPR pairs. The parameters are $L=3, \eta=p_1=1$, and $p_2=0.97$. The (upper) purification curve corresponds to scheme A, Eq.~(\ref{modified_bennett}). The (lower) connection curve is described by (\ref{connect})
with $N=3$.}
\label{FIGpuriloop}
\end{figure}

The central feature of this nested purification protocol is the two-step process connection - purification on each nesting level. This purification loop is visualized in Fig.~\ref{FIGpuriloop}. The curves shown in this figure are described by the formulas
(\ref{modified_bennett}) for purification (upper curve), and (\ref{connect}) for connection (lower curve, with $N$ replaced by $L$), respectively.
Let us consider a given nesting level k, where we have $N/L^{k-1}$ pairs of fidelity F each. Starting from F, the fidelity $F_L$ after connecting $L$ pairs can be read off from the curve below the diagonal. Reflecting this value back to to the diagonal line, as indicated by the arrows in Fig.~\ref{FIGpuriloop}, sets the starting value for the purification curve. If $F_L$ lies within the purification interval $(F_{\rm min},F_{\rm max})$, then the purification curve lies above the diagonal, $F'(F_L)>F_L$, and iterative application of (\ref{modified_bennett}) leads back to a fidelity larger or equal to the initial value $F$. 
Each step in the `staircase' of Fig.\ref{FIGpuriloop} corresponds to a successful purification step, and with help of the total number of necessary steps $k_{max}$ needed to recover one can calculate the needed resources M as described in Section \ref{Purification}. Once the initial value $F$ of the fidelity is reobtained, we have $N/L^k$ pairs and we can start with the next level k+1. In summary, each level in the protocol corresponds to one cycle in Fig.~\ref{FIGpuriloop}. 

There are two conditions that have to be fulfilled to realize a closed loop. First, the fidelity after the connection has to be larger than the minimum value for purification (lower fixpoint), $F_L > F_{\rm min}$. If $F_L\le F_{\rm min}$, then repurification is impossible. Second, the fidelity $F$ one starts with and wants to reach again has to be smaller than the maximal reachable fidelity (upper fixpoint), $F < F_{\rm max}$. These two conditions determine both a threshold value for the error tolerances
of the local operations and they limit the value $L$ (the number of pairs which can be connected before purification becomes necessary). Please note that, being polynomial in $L$, the lower curve gets steeper and steeper near $F=1$ for higher and higher values of $L$. From this one can see that, for a given initial fidelity $F$, there is a maximum number of pairs one can connect before repurification becomes impossible.

The threshold value for the nested purification scheme, i.e. the repeater, is tighter than the threshold for simple purification on a single segment. Consider, as an example, the situation in Fig.~\ref{FIGpuriloop}. The chosen value $p_2=0.97$ of the error parameter 
(with $\eta=1$) lies above the purification threshold of $0.95$ that we have found after Eq.~(\ref{simple_fixpoints}). On the other hand, for $p_2=0.95$, the purification interval
in Fig.~\ref{FIGpuriloop} would shrink essentially to a point at $F_{\rm min}=F_{\rm max}=
0.75$. Clearly, in this situation no loop can be realized; even if we start initially
at $F=0.75$, the purification ``interval" is left after the first connection process
and the fidelity would subsequently converge towards the trivial fixpoint $F=0.25$.

When we employ the protocols A or B in the purification part of the nested scheme, 
we find error thresholds that are typically in the percent region. 
For scheme C, the situation is similar, {\em as long as} we do not use depolarization after each purification step (i.e. keep all involved pairs during the purification process in the Werner form). If we do use depolarization, thus realizing a modified version of protocol A,
the loop cannot be closed.  Differently speaking, for Werner states, the fidelity that is lost by the connection process can never be regained by purification with this scheme.
For this to be true, the repurification condition $F_{\rm max}(F_L(F))\ge F$ must be fulfilled, where the upper fixpoint $F_{\rm max}$ is a function of the fidelity of the
auxiliary $L$-pair. It can be shown analytically that this is not possible for scheme C with Werner states. This underlines the fact that the purification scheme C is not a trivial variant of either of the schemes A and B. 

Note that both connection and purification are not smooth processes. This means that one can only perform whole steps and not parts of a step. Due to this fact, one will
not get an exactly closed purification loop, in general, but the final fidelity after the purification (this is the value at which the ladder in Fig.~\ref{FIGpuriloop} ends)  may be slightly larger than the initial fidelity; secondly, the final fidelity may be different at each nesting level, but it is always larger or equal to the fidelity of the elementary pairs.


\subsection{Physical resources}\label{Properties}

The length of a segment is limited by the transmission errors, and the number of segments $N$ one splits up the total channel should be varied in order to find out the optimal configuration. 
The quantum repeater can be characterized by two quantities, which depend on the used connection and purification protocol
\begin{itemize}
\item The total {\it physical resources per segment} needed to build up the EPR pair - this gives the number of necessary 'parallel channels' between two checkpoints.
\item The {\it total time} which is needed to build up the EPR pair. 
\end{itemize}       
The way to calculate these quantities is similar for schemes A and B, while the properties of scheme C are totally different.

\subsection{Physical resources per segment}
The physical resources per segment $R_{segment}$ gives us the number of required connection lines between two checkpoints, and also half of the particles which should be stored at each checkpoint. On the borders (at A and B), it is equal to the number of particles to be stored, while at each checkpoint $C_i$, one has to take care about the incoming and outgoing string and therefore needs the double number of particles. $R_{segment}$ (vertical axis in Fig.\ref{FIGnesting}) is completely determined by the resources M needed for each purification loop and is given by $R_{segment}=M^n$, where $n=log_{L}N$ is the number of nesting levels. The resources $M$ needed for purification can be calculated as described in section \ref{Purification}.

\begin{figure}
\begin{picture}(230,200)
\put(0,0){\epsfxsize=230pt\epsffile[24 180 540 591 ]{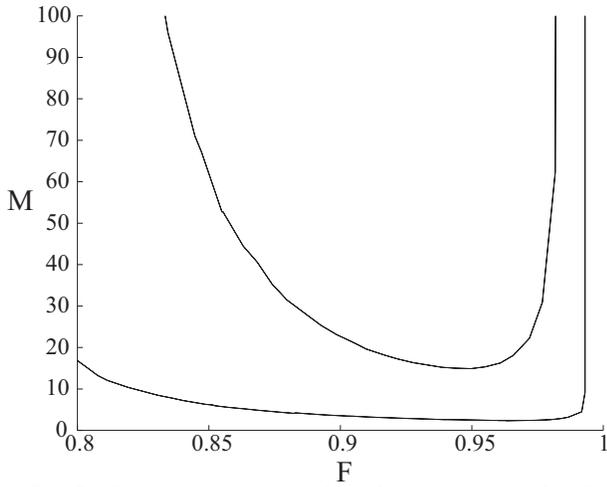}}
\end{picture}
\caption[]{Resources $M$ for purification versus working fidelity $F$ for scheme A (upper curve) and the scheme B (lower curve). 
The errors of all operations are 0.5\%, and $L=2$.}
\label{FIGbennett}
\end{figure}

For schemes A and B, the physical resources are equal to the resources, since all pairs have to be created and therefore stored at the beginning of the process. Figure \ref{FIGbennett} shows the resources $M$ needed for a single purification loop, where it is assumed that $L=2$, i.e. only two pairs are connected before re-purification takes place. $M$ is plotted against a 'working fidelity', which is the fidelity one starts with before connection and one ends up after re-purification. Although the difference between schemes A and B does not look very dramatic,  note that $M$ has to be taken to the $n^{th}$ power to calculate the resources per segment, where for long range communication $n\approx10$ (see Sec. \ref{Compare}). One can see that there exists an optimal working region, which can be understood as follows: 
For small working fidelities F, two things happen: Firstly due to the polynomial law for the connection, the fidelity decreases strongly. Secondly, purification becomes less effective near the lower fixpoint $F_{\rm min}$ (the gain per purification step gets smaller). For both reasons, many purification steps are needed to recover and thus the needed resources are large.
For working fidelities F close to the maximal reachable fidelity (upper fixpoint $F_{\rm max}$), one does not loose much due to the connection, but purification is less effective - the gain per purification step is smaller. Therefore the the number of necessary steps is again quite large and so are the needed resources.
In between there exists a region which is optimal in the sense that the needed resources are minimized. Here a working fidelity $F\approx 0.95$ turns out to be optimal for the chosen error parameters.  

\begin{figure}
\begin{picture}(230,200)
\put(0,0){\epsfxsize=230pt\epsffile[40 176 547 592]{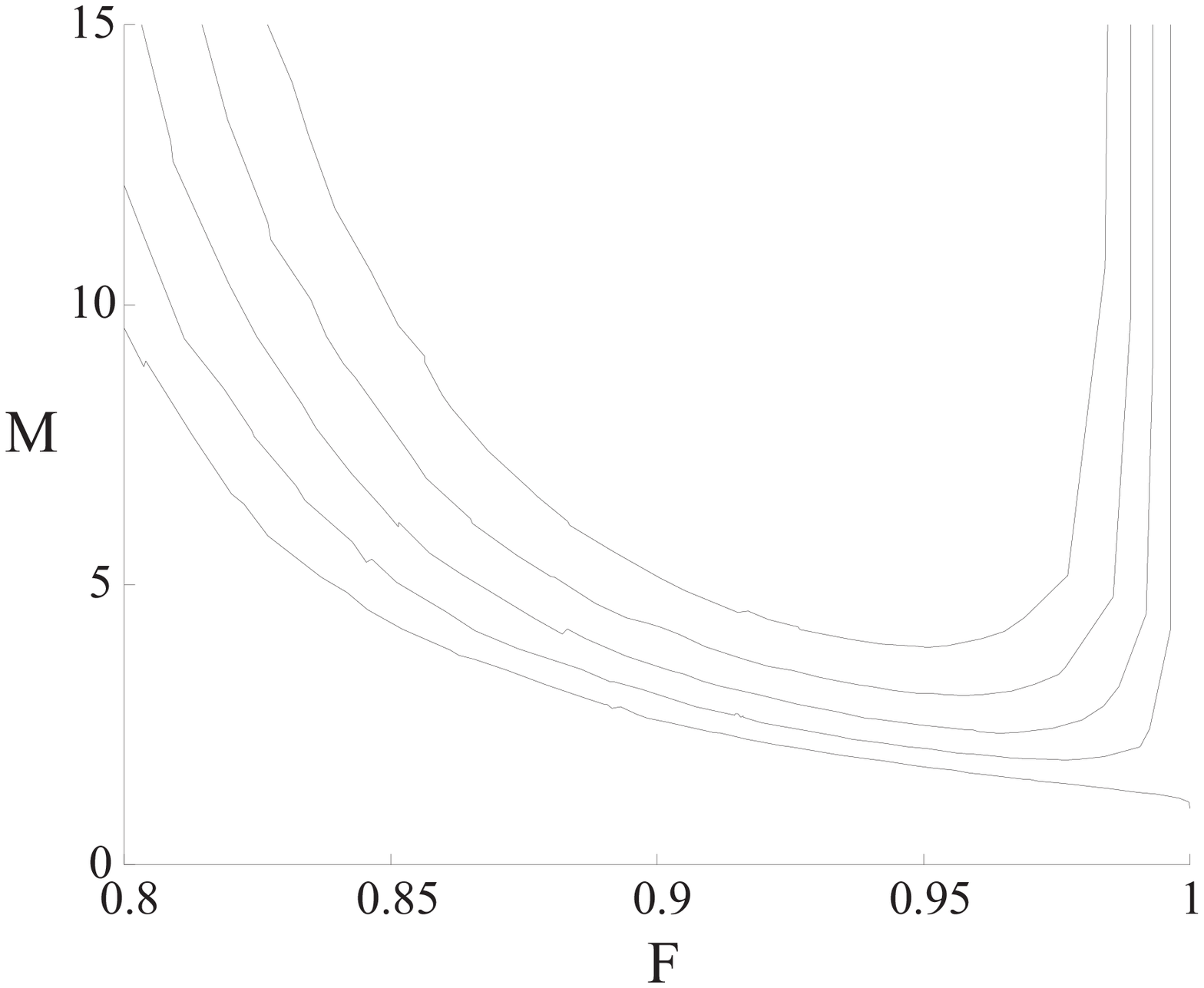}}
\end{picture}
\caption[]{Resources $M$ versus working fidelity $F$ for different error 
parameters (scheme B with $L=2$). The curves from bottom to top correspond to error 
probabilities of $0\%$, $0.25\%$, $0.5\%$, $0.75\%$, and $1\%$. The center curve correspond 
to the lower curve in Fig.~\ref{FIGbennett}.}
\label{FIGoxford}
\end{figure}

The error dependence of the resources is shown in Fig.\ref{FIGoxford} . For smaller error probabilities (larger error parameters), two things happen: The optimal working fidelity gets larger and the needed resources $M$ decrease. For imperfect local operations, the needed resources increase to infinity if one really wants to reach the maximal achievable fidelity $F_{\rm max}$, only for perfect operations and $F=1$, this is not true because one starts with perfect pairs and does not loose due to connection. Therefore no purification is necessary and $M=1$.

Using scheme C for purification, the situation is different.  
The vertical axis of Fig.\ref{FIGnesting} (which corresponds to the resources) is translated into a "temporal axis". Instead of creating all needed pairs at the beginning and operating parallelly on ensembles of pairs at each checkpoint (as in scheme A and B), one creates the pairs needed for purification every time one needs one and operates sequentially on the pairs. To perform the purification process, every time only two pairs are involved. One is needed to store the purified state and the other one is repeatedly created to purify the first pair as visualized in Fig.\ref{IbkResfig}. The number of {\it physical resources} required for purification is thus only 2, but every time the purification was not successful one has to start from the very beginning. Using scheme A or B one obtains a purified pair after $k_{max}$ successful purification steps, which is in this case also the number of really performed purification steps. Looking at Fig.\ref{tree}, one sees that a purification step which was not successful eliminates one branch of the tree, but after 4 steps (in this example) one ends up with a purified pair. In scheme C, on the other hand, one does not follow all branches parallel but follows the branches sequentially (and using always the same elementary pair for each purification step). This limits the number of involved particles, but increases the required time.

\begin{figure}
\begin{picture}(230,170)
\put(0,0){\epsfxsize=440pt\epsffile[0 0 330 68]{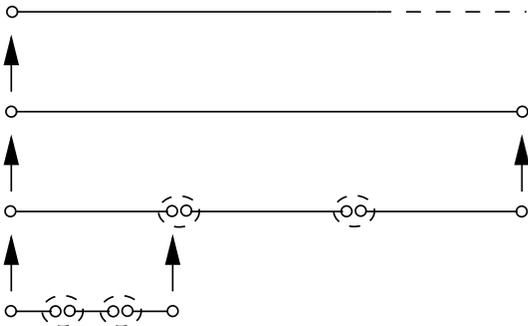}}
\end{picture}
\caption[]{The nested purification algorithm using the Innsbruck protocol. At each nesting level, one additional particle at the borders of this nesting level is needed to store the pair meanwhile}
\label{IbkRep}
\end{figure}

Looking now at the nested algorithm, it turns out that the required number of {physical resources} for scheme C grows linearly with the number of nesting levels and thus logarithmically with the number of segments (and the distance). This is because one additional pair is needed for storage purposes at each nesting level. One can understand this by inspecting Fig.\ref{IbkRep}. First, three elementary pairs are connected (line 4) and used to purify the pair at line 3, which now is the 'elementary pair' for the next nesting level. Second, three of these re-purified pairs (line 3) are connected and used to purify the pair at line 2. Since one has to create such a pair repeatedly and all particles at lines 3 and 4 are involved, it is necessary to store the pair to be purified (here in line 2).   
Therefore one needs one additional particle to store the pair at each new nesting level. This particle is not needed in all checkpoints, but only in those which lie on the {\it borders} of the corresponding nesting levels. That is why the maximum number of additional particles (physical resources) is required at the outermost places (A and B), while e.g. at checkpoint $C_1$, no additional particle is needed. The {\it maximal} number of additional particles (physical resources) grows only linearly with the number of nesting levels and therefore logarithmically with the number of segments in contrast to the polynomially growing resources when using scheme A or B. The maximal resources per segment (in this case particles needed to store the pair) can be calculated by 
\be R_{segment}=1+log_LN=1+n \ee

\subsection{Temporal resources}

Here we will discuss the total time needed to create an EPR pair. This involves at least three parameters:
\begin{itemize}
\item The typical time $\tau_{op}$ needed to perform local operations (single qubit, two qubit and measurements).
\item The typical time $\tau_{pair}$ to create an elementary pair. Using optical fibers and the model of the absorption free channel (AFC) \cite{vanenk97b,vanenk97a}, this can be expressed in terms of the other 2 parameters and is given by 
\be \tau_{pair}=\tau_{AFC}=\left(5\tau_{op}+2\tau_{class}\right)e^{\left(\frac{l_{segment}}{l_0}\right)} \ee
where $l_0$ is the half length of the fiber and $l_{segment}$ is the length of a segment. This just reflects the fact that for a single use of the AFC, all together 5 operations and 2 classical communications (in this case the transmission of a photon) are necessary. The exponential function gives the average number of repetitions which are necessary due to absorption losses. 
\item The time $\tau_{class}$ needed for classical communication to broadcast the result of the measurement. This time is determined by the length of the segment and the speed of light c and is given by 
\be \tau_{class}=\frac{l_{segment}}{c} \ee
\end{itemize}
To simplify the discussion, we will assume that the number of pairs $L$ which are connected at each nesting level is some power of 2, $L=2^l$, because in this case the connection process can be performed in parallel. For each connection process, 3 elementary operations (CNOT, measurement, operation depending on the outcome of the measurement) and classical communication over the distance of one segment are needed. For purification, also three elementary operations (e.g. for scheme A depolarization, bilateral CNOT and bilateral measurement) and classical communication of the result are necessary.

\begin{figure}[ht]
\begin{picture}(230,210)
\put(0,0){\epsfxsize=230pt\epsffile[0 0 321 297]{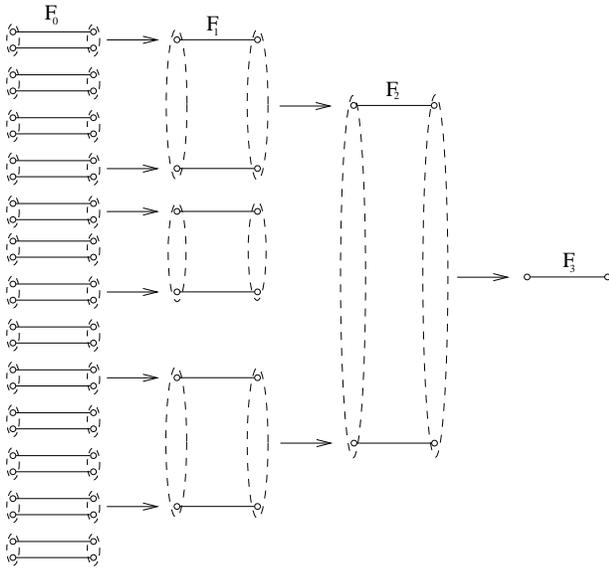}}
\end{picture}
\caption[]{Typical purification process. At each purification step, either both pairs are discarded (if the purification was not successful) or one pair is discarded (if the purification was successful). The left-over pairs are again used for purification at the next step.}
\label{tree}
\end{figure}

Using scheme A or B, the time needed to perform a {\it purification loop} (the connection of $L=2^l$ pairs and re-purification, where $k_{max}$ successful purification steps are necessary) at nesting level m is given by

\be t_{loop}(m)= 3l\tau_{op}+f(m)(2^k-1)\tau_{class}+k_{max}(3\tau_{op}+f(m)2^l\tau_{class}) \ee  

where $f(m)=\left(2^l\right)^m$ gives the length dependence of the elementary pair on the nesting level. The first two terms give the time required to perform the connection process in parallel fashion, where $l$ connections and classical communication over the distance of $(2^l-1)$ times the length of an elementary pair is needed. The purification process involves $k_{max}$ purification steps and for each purification step classical communication over the distance of $2^l$ times the length of an elementary pair (since $L=2^l$ pairs were connected before). One only has to take into account the successfull purification steps, since one runs through the whole purification tree (see Fig.\ref{tree}) in a parallel fashion.
To calculate the total time needed to perform the nested algorithm, one can use the iteration formula
\be t_{tot}(m)=t_{tot}(m-1)+t_{loop}(m) \label{Iteratetime}\ee
where m corresponds to the $m^{th}$ nesting level and $t_{tot}(n)$ gives the total time needed for the creation of an EPR pair between A and B. To complete this iteration formula, note that $t_{tot}(0)=\tau_{pair}$ which is just the time needed to create an elementary pair at the beginning. The total time if one has $N=L^n$ segments and $L=2^l$ is thus
\be  t_{tot}^{A,B}= n(3l+3k_{max})\tau_{op}+\left(2^l-1+k_{max}2^l\left(\frac{(2^l)^n-1}{2^l-1}\right)\right)\tau_{class} +\tau_{pair} \ee
The main contribution is given by the classical communication if $\tau_{op}$ is not too large and is just $t_{tot}\approx(2k_{max}+1)N\tau_{class}$.

Using scheme C for purification, the time needed to perform a purification loop on nesting level m is given by
\be t_{loop}(m)=\left(\text{max}\{t_{loop}(m-1)\}+3l\tau_{op}+f(m)(2^l-1)\tau_{class}\right)M +S(3\tau_{op}+f(m)2^l\tau_{class}) \ee
where the first term is the time needed to create a pair for purification (by building up $2^l$ pairs at nesting level (m-1) and connecting them), which must be done {\it sequentially} M times. The second term gives the time needed for purification (operations and classical communication over the distance of $2^l$ times the length of an elementary pair), where S purification steps have to be performed.  The function $f(m)=\left(2^l\right)^m$ reflects the length dependence of the elementary pair on the nesting level and the time needed to create an elementary pair at the lowest nesting level $\tau_{pair}=t_{loop}(0)$, which completes this iteration formula. 
The maximum appearing in this expression is taken over all $2^l$ segments on nesting level $(m-1)$. This is necessary because all $2^l$ new elementary pairs have to be built up before the next connection process can start. Therefore one has to wait for the slowest one. Note that the time for purification is fixed for schemes A and B (obtaining bad results in this case corresponds to the elimination of a branch in Fig.\ref{tree}), while for scheme C the purification process has to be started from the very beginning when obtaining a bad result. This results in the fact that the performed purification steps S and the used purification resources M (and therefore the needed time) might vary from case to case. This formula is sufficient to calculate the total time needed to perform the nested algorithm up to nesting level m and $t_{tot}(n)=t_{loop}(n)$ gives the total time needed to build up an EPR pair between A and B. 

A rough estimation of the total time can be obtained when using (\ref{iteration}) and (\ref{iteration1}) to calculate the {\it average} number of purification resources M and purification steps S. In this case, one can write down a closed expression for the total time, which is polynomially growing with the number of segments $N$ and thus with the distance. Comparing this time with the results obtained from a simulation of the whole process, it turns out to be smaller by a factor of $\approx3$.  
 In some sense, the vertical axis (physical resources) in Fig.\ref{FIGnesting} is translated into a temporal axis.

\subsection{Comparison of the 3 schemes}\label{Compare}
 
We will give some typical numbers to quantify the properties of the nested purification protocol if the purification loop is performed with help of scheme A,B or C. The following tables are based on error parameters of $\frac{1}{2}\%  (p_1=p_2=\eta=0.995)$ and a working fidelity F=0.96, the fidelity one starts with at the lowest nesting level and one ends up after each purification loop.  
The length of a segment is assumed to be in the order of half length of an optical fiber, $l_{segment}\approx 10km$. In this case , the number of segments is proportional to the distance between two locations. If the number of segments is in the order of $N=2^7=128$, this will be called 'Continental scale' (distance of $\approx 1000km$), while $N=2^{10}=1024$ is called 'Intercontinental scale' (distance of $\approx 10000km$), since one can create an EPR pair between two international cities.
It is assumed that the connection is performed in parallel. 
The columns from left to right have the following meaning:

\begin{itemize}
\item resources: this gives the {\it physical} resources per segment.
\item time: gives the total time in seconds needed to create an EPR pair. The number is based on the application of the AFC to create elementary pairs ($\tau_{pair}=\tau_{AFC}=3.2*10^{-4}s$). The half length of the fiber is considered to be 10km, which is also the length of the elementary segment ($\tau_{class}=0.33*10^{-4}s$). All operations (single qubit, 2 qubit, measurement) are considered to be performed in $\tau_{op}=10^{-5}s$. The time is calculated using a simulation of the nested entanglement purification process, averaging over a few hundred runs. 
\end{itemize}

\begin{tabular}{|l|c|c||c|c|} \hline
 & \multicolumn{2}{c||}{Continental scale} & \multicolumn{2}{c|}{Intercontinental scale} \\ \hline
& resources &time & resources &time  \\ \hline
A& $1.58*10^9$ & $3.88*10^{-2}$ & $9.01*10^{12}$ & 0.298 \\ \hline
B& 329 &  $1.34*10^{-2}$ & 4118 & 0.103 \\ \hline
C& 7 &  0.77 & 10 &15.69\\ \hline
\end{tabular}
\vspace{0.5cm}

One sees that the physical resources needed when using scheme C are some orders of magnitude smaller, while the time needed is larger. For a practical implementation, the achievable bit rates are not very impressive, but note that all numbers get much better if the local operations can be performed better (error parameters closer to 1) as can be seen in Fig.\ref{FIGoxford}. 

So the reachable fidelity is larger and at the same time, the required resources are smaller when using better local operations.
Note that the total time needed is in the order of the time for classical communication over this distance, which is just $t_{class}=\frac{10240km}{3*10^5km/s}=0.034s$.


\section{Summary}

We have shown that it is possible to create EPR pairs via a noisy channel, 
with an overhead in physical resources that grows only logarithmically with the 
length of the channel, while the time needed for the creation grows polynomially. 
The central idea is to use a nested purification 
protocol for connecting a sequence of EPR pairs at certain ``connection points" within the 
channel, whose roles are reminiscent of to classical repeaters. Different from the 
classical situation, the concept of the quantum repeater is not a local amplifier, but
it involves both the local checkpoints and global purification protocol. 
Our scheme tolerates errors for local operations and measurements that are 
in the percent region. 

This work was supported in part by the \"Oster\-reich\-isch\-er Fonds zur F\"or\-der\-ung
der wis\-sen\-schaft\-lich\-en Forschung, and by the European TMR network 
ERB-FMRX-CT96-0087.




\begin{figure}[ht]
\begin{picture}(230,79)
\put(0,0){\epsfxsize=230pt\epsffile[0 0 456 79]{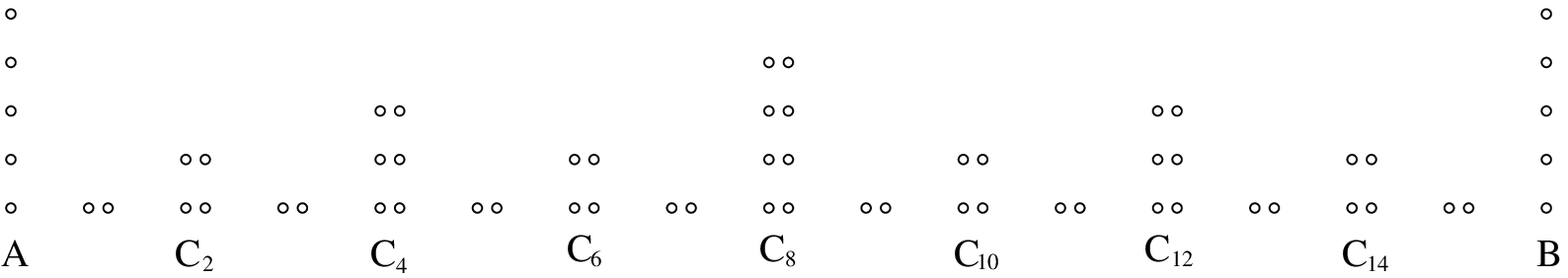}}
\end{picture}
\caption[]{Additional particles needed at each checkpoint for the Innsbruck protocol. The number of segments $N=2^4=16$, L=2 pairs are connected at each nesting level and the number of nesting levels is n=4.} 
\label{Ibkparticles}
\end{figure}


\begin{references}
\bibitem[*]{onleave} Permanent address: Institut f\"ur Theoretische Physik, 
    Universit\"at M\"unchen, Theresienstrasse 37, D--80333 M\"unchen.
\bibitem{schumacher96}
    B. Schumacher, Phys. Rev. A, {\bf 54}, 2614 (1996); B. Schumacher and 
    M. D. Westmoreland, {\em ibid.} {\bf 56}, 131 (1997).
\bibitem{wootters82}
    W. K. Wootters and W. H. Zurek, {\it Nature} {\bf 299}, 802 (1982).
\bibitem{dieks82}
    D. Dieks,  Phys. Lett. A {\bf 92} 271 (1982). 
\bibitem{glauber86}
    R. J. Glauber, in {\em Frontiers in Quantum Optics}, 
    (eds. E. R. Pike \& S. Sarkar), pp.\ 534-582. Bristol: Adam Hilger.
\bibitem{tittel97}
    W. Tittel, J. Brendel, B. Gisin, T. Herzog, H. Zbinden, and N. Gisin, 
    {\it Los Alamos report} quant-ph/9707042.
\bibitem{footnote0}
    For optical fibers, this length is typically 10km; see e.g.\ \cite{tittel97}. 
\bibitem{fault_tolerant}
    P. Shor, quant-ph/9605011; A. M. Steane {\it Phys. Rev. Lett.} 
    {\bf 78} 2252 (1997). D. Gottesman, quant-ph/9702029; 
    For a review see e.g. J. Preskill, quant-ph/9712048.
\bibitem{knill96}
    E. Knill and R. Laflamme, quant/ph-9608012.
    See also D. Aharonov and M. Ben-Or, quant-ph/9611025; C. Zalka, 
    quant-ph/ 9612028.
\bibitem{bennett96a}
    C. H. Bennett, G. Brassard, S. Popescu, B. Schumacher, J. A. Smolin,
    and W. K. Wootters, Phys. Rev. Lett. {\bf 76}, 722 (1996).
\bibitem{bennett96b}
    C. H. Bennett, D. V. DiVincenzo, J. A. Smolin, and W. K. Wootters 
    {\it Phys. Rev. A} {\bf 54}, 3824 (1996).
\bibitem{gisin96}
    N. Gisin, Phys. Lett. A {\bf 210} 151 (1996).
\bibitem{deutsch96}
    D. Deutsch, A. Ekert, C. Macchiavello, S. Popescu, and A. Sanpera, 
    Phys. Rev. Lett. {\bf 77 }, 2818 (1996).
\bibitem{bennett93}
    C. H. Bennett, C. H. Brassard, C. Crepeau, R. Josza, A. Peres, and W. K. Wootters,
    Phys. Rev. Lett. {\bf 70}, 1895 (1993).
\bibitem{ekert91}
    A. K. Ekert, {\it Phys.\ Rev.\ Lett.} {\bf 70}, 661-663 (1991).
\bibitem{giedke98}
    G. Giedke {\em et al.}, unpublished. 
\bibitem{briegel98}
    H. J. Briegel, W. D\"ur, J. I. Cirac, and P. Zoller, quant-ph-9803056 (1998).
\bibitem{zurek96}
    C. Miquel, J. P. Paz and W.H. Zurek,    Phys. Rev. Lett. {\bf 78}, 3971 (1997).
\bibitem{footnote1}
    A similar form as in (\ref{result_bloch_av}) may be obtained for two-bit             rotations with a corresponding isotropic noise function, although the calculation is more         complicated. In particular, we have not yet been successful in giving a proof in terms of a     manifest {\em parametrization-independent average} over the invariant group manifold of SU(4). Isotropic noise in this case means that the depolarizing channel does not 
distinguish between the two qubits and acts on them in a symmetric way.
\bibitem{footnote2}
    This assumption is not essential; it is motivated by the fact
    that the depolarization is a itself an inessential although 
    simplifying step in the Bennett et al. protocol. 
    If the reliability $p_1$ is the same for all one-bit operations involved in the                                              realization of the twirl, an {\em imperfect depolarization} can effectively be
    described by
    \[
       T \rho_{12}' = p_T T^{\text id} \rho_{12}' + \frac{1-p_T}{2} I_{12} 
    \] 
    where $p_T=(p_1)^2$ summarizes reliability of the total twirl operation $T$ and
    $\rho_{12}'$ is the bilocal state before the twirl is applied.
    This operation still depolarizes $\rho_{12}'$ and  brings it to the
    Werner form. However, the fidelity $F''$ of the resulting state is reduced
    with respect to the fidelity $F'$ of the state $\rho_{12}'$ in (\ref{modified_bennett})
    by the relation 
    \[
     F'' = p_T F' + \frac{1-p_T}{4}\,. 
    \]
\bibitem{helstrom76}
    C. W. Helstrom, {\it Quantum detection and estimation theory}, Associate Press, 
    London 1976.
\bibitem{composed_op}
    Any sequence of operations at the same node, e.g. two single qubit operations             $O_1\otimes O_2$ followed by a CNOT operation will be described by a single two-qubit         operation $O_{12}$ and consequently with only a single error parameter
\bibitem{Duer98}
    W. D\"ur, Master Thesis, available under http://www.uibk.ac.at/c/c7/c705/staff/wd.html 
\bibitem{Luetkenhaus}
    see also N. L\"utkenhaus, unpublished. 
\bibitem{entanglement_swapping}
    A. Zeilinger {\em et al.}, {\it Phys. Rev. Lett.}
    {\em 78}, 3031 (1997)  
    In their scheme, however, a three-particle entangled state is obtained rather than a distant EPR pair. 
\bibitem{footnote3}
    This procedure is not necessary, but to have a consistent model for both connection
    and purification, we want to describe all operations in terms of 
    (\ref{two_qubit_op}--\ref{povm}). 
\bibitem{vanenk97b}
    S. J. van Enk, J. I. Cirac, and P. Zoller, {\it Science}, {\bf 279}, 205 (1998).
\bibitem{vanenk97a}
    S. J. van Enk, J. I. Cirac, and P. Zoller, {\it Phys. Rev. Lett.}, {\bf 78}, 4293 
    (1997).
\end{references}
\end{document}